\documentclass[aasms,tighten,psfig,graphicx,11pt]{aastex}
\slugcomment{Submitted to {\it AJ.}, accepted 2000}

\begin{document}

\title{Population and Size Distribution of \\
        Small Jovian Trojan Asteroids}

\author{David C.\ Jewitt, Chadwick A.\ Trujillo}
\affil{Institute for Astronomy, University of Hawaii,\\
2680 Woodlawn Drive, Honolulu, HI 96822}

\and

\author{Jane X.\ Luu}
\affil{Sterrewacht Leiden\\
Postbus 9513, 2300RA Leiden, The Netherlands}

\newpage

\begin{abstract}
	
We present a study of Jovian Trojan objects detected serendipitously
during the course of a sky survey conducted at the University of
Hawaii 2.2-meter telescope.  We used a 8192$\times$8192 pixel
charge-coupled device (CCD) mosaic to observe 20 deg$^2$ at locations
spread over the L4 Lagrangian swarm and reached a limiting magnitude
$V = 22.5$~mag (50\% of maximum detection efficiency).  Ninety-three
Jovian Trojans were detected with radii $2 \leq r \leq 20$ km (assumed
albedo 0.04).  Their differential magnitude distribution has a slope
of $0.40 \pm 0.05$ corresponding to a power law size distribution
index $3.0 \pm 0.3$ ($1\sigma$).  The total number of L4 Trojans with
radii $\geq 1$ km is of order $1.6 \times 10^5$ and their combined
mass (dominated by the largest objects) is $\sim$$10^{-4}~M_{\rm
Earth}$. The bias-corrected mean inclination is $13.7^{\circ} \pm
0.5^{\circ}$.  We also discuss the size and spatial distribution of
the L4 swarm.

\end{abstract}


\newpage

\section{Introduction}

The Jovian Trojans are asteroidal objects confined to two swarms in
Jupiter's orbit, leading and trailing the planet by $60^{\circ}$ of
longitude (known as the L4 and L5 Trojans, respectively).  The first
recognized Jovian Trojan (588 Achilles), discovered in 1906 by Max
Wolf, was taken as providing observational confirmation of Lagrange's
prediction of stable orbits at the triangular points.  Currently, 132
Jovian Trojans have been numbered while another 125 await permanent
designations.  These objects follow loose orbits that librate around
the L4 and L5 points with periods near 150 years.  Recent work has
shown that Trojan orbits are destabilized by collisional ejection (for
which the loss rate of bodies larger than 1 km in diameter is
estimated at $\sim$$10^{-3}$~yr$^{-1}$; Marzari et al. 1997) and, to a lesser
extent, by dynamical chaos (corresponding loss rate $\sim$$6 \times
10^{-5}$~yr$^{-1}$; Levison, Shoemaker and Shoemaker 1997).  The
implication is that the Trojans must either be the remnants of a much more
substantial initial population of trapped bodies or that these objects
are continually replenished from an unidentified external source.

The origin of the Trojans is a subject of much conjecture.  The
principal dynamical problem concerns the nature of the dissipation
needed to stabilize objects in weakly bound orbits librating about L4
and L5.  Schemes under consideration include capture of near-Jupiter
planetesimals by gas drag in an early phase of the solar nebula (Peale
1993), stabilization of planetesimals near the L4 and L5 points
due to the rapidly increasing mass of Jupiter in the late stages of
its growth (Marzari and Scholl 1998), and collisional dissipation
followed by capture of asteroidal fragments (Shoemaker et al. 1989).
Physical observations provide only limited clues about the source of
the Trojans.  The optical (Jewitt and Luu 1990; Fitzsimmons et
al. 1994) and near-infrared (Luu, Jewitt and Cloutis 1994; Dumas, Owen
and Barruci 1998) reflection spectra appear featureless, and are
reminiscent of the spectra of the nuclei of short-period comets.  Like
cometary nuclei, the Trojans have very low ($\sim$4\%) visual albedos
(Cruikshank 1977; Tedesco 1989) that suggest carbonized surface
compositions.  If the Trojans formed near or beyond Jupiter's orbit,
temperatures were probably low enough for water to exist as solid ice
(rather than vapor, as in the inner nebula).  This fact has led to the
suggestion that the Trojans might possess ice-rich interiors
equivalent to those of the cometary nuclei, a possibility which is not
contradicted by any available observations (Jewitt 1996).

In this paper, we discuss the results of an optical survey taken in
the direction of the L4 Jovian swarm.  The survey differs from most
previous work on these objects in two main respects.  First, it is
based on the use of a digital (CCD) detector instead of photographic
plates and so has relatively high sensitivity to faint (small) Jovian
Trojans.  Second, the parameters of the survey are extremely well
known as a consequence of the relative ease with which digital data
may be calibrated (compared to non-linear, analog photographic data).
Therefore, we are able to measure the statistical properties of the L4
Trojans with greater confidence than would be possible with
photographic data.  A preliminary abstract describing this work (Chen
et al. 1997) is superceded by the present report.

\section{Observations and Data Reduction}
	
The present observations were taken as part of a study of the Kuiper
Belt, the main results of which are already published (Jewitt, Luu and
Trujillo 1998).  Here we present observations taken UT 1996 Oct.\ 7 to
15 at the $f/10$ Cassegrain focus of the University of Hawaii
2.2-meter telescope with a 8192$\times$8192 pixel CCD mosaic
(hereafter called 8k).  The 8k consists of eight 2048$\times$4096
pixel Loral chips with 15 $\mu$m pixels and gaps between chips of
$\sim$1~mm. The pixels were binned 3$\times$3 in order to reduce the
readout time (from approximately 6 minutes to 1 minute) while
maintaining Nyquist sampling of the images.  The binned image scale
was $0.405 \pm 0.002~''$/pixel yielding a field of view
18.4'$\times$18.4' (0.094 deg$^2$).  Typical image quality (including
contributions from the intrinsic seeing, wind shake and tracking
oscillations during the unguided integrations) varied from 0.8'' to
1.0'' Full Width at Half Maximum (FWHM), meaning that the images were
Nyquist sampled.  The images were taken with a 150-sec integration
time through a specially optimized VR filter (bandwidth 5000 \AA~to
7000 \AA; Jewitt, Luu and Trujillo 1998).  Each sky position was
imaged at three epochs, with a separation between epochs of about 1
hour.  In total, we observed 20 deg$^2$ of sky in the direction of L4.

The data were flattened using a median combination of dithered images
of the evening twilight sky.  Observations of photometric standard
stars (Landolt 1992) were used to calibrate the sensitivity of each
chip.  By defining an A0 star to have $m_{VR} = V = R = 0$, an object of
solar color ($V - R = 0.35$) has $V \approx m_{VR} + 0.2$.  We adopted the
latter relation to transform our VR magnitude to standard $V$ magnitude
in this work.  Note that Trojan asteroids display a wide range of
optical colors, from nearly solar to very red ($V - R \sim 0.6$: Jewitt
and Luu 1990, Fitzsimmons et al. 1994), leading to the introduction of
small, color-dependent corrections to the $V$ vs. $m_{VR}$ relation.  In
addition, some of the 8k CCDs were of locally inferior photometric
quality.  Together, these effects introduce an inherent uncertainty in
the absolute photometric accuracy of about 0.2 mag.  Trojans were
identified using the MODS detection program (Trujillo and Jewitt
1998).  We determined the detection efficiency of MODS by searching
for artificial objects added to real data.  The efficiency is
adequately fitted by the function

\begin{equation}
e = e_{\rm max} \left[1 - \frac{1}{2}\exp \left(\frac{m_V - m_V(50)}{\sigma_V} 
\right) \right]
\end{equation}

\noindent where $e_{\rm max}$ is the maximum detection efficiency, $m_V$
is the Trojan magnitude, $m_V(50)$ is the magnitude at which the
detection efficiency equals $e_{\rm max}/2$, and $\sigma_V$ measures the
width of the band of decreasing detection efficiency.  When including
the trailing loss due to the motion of the Trojans during the
integration, we found $e_{\rm max} = 0.86 \pm 0.01$, $m_V(50) = 22.47 \pm
0.05$ and $\sigma_V = 1.13 \pm 0.01$.  Variations in the seeing within
the data are small and affect $m_V(50)$ by at most $\sim 0.1$
mag. (i.e., less than the formal uncertainty in the absolute
photometry).

We took all observations near opposition, where the rate of retrograde
motion across the sky, $\omega$ [$''$/hr], is inversely related to the
heliocentric distance.  With this constraint we could not observe the
L4 point directly, but instead mapped areas at a range of angular
distances from the Lagrangian point (Figure~1).  In addition, because
we could not afford to interfere with our primary (Kuiper Belt)
observational program, we secured no follow-up astrometry of Trojan
candidates with which to determine orbital elements.  Instead, we used
the sky-plane angular speed to distinguish Trojans from main-belt
asteroids.  The main-belt asteroids move westward at rates generally
higher than the more distant Trojans, permitting us to separate the
two types of object on the basis of speed.  Figure 2a shows the
apparent velocities in RA and Dec of numbered main-belt asteroids and
previously known Trojans in the direction of our observations on 1996
October 10.  The curves in Figure~2a show the loci of points having
total angular motions $\omega = 22''$/hr and $\omega = 24''$/hr.  At $\omega >
24''$/hr we find exclusively main-belt asteroids (Fig. 2a).
Roughly equal numbers of main-belt and Trojan asteroids appear in the
range $22 \leq \omega \leq 24''$/hr.  On the other hand, a large
majority ($\sim$90\%) of the objects with $\omega \leq 22''$/hr are
Trojans. Therefore, $\omega \leq 22''$/hr constitutes our operational
definition of Trojans in this survey.  This definition is clearly not
perfect, but it is sufficiently robust that we can make statistical
identifications of the Jovian Trojans in our data.  Figure~2b shows
all 93 Trojan candidates flagged by the detection software.  Of these,
only the 4th object in Table 1 has position and motion consistent with
a previously known Trojan asteroid (6020 P-L).  The best evidence that
we have indeed obtained a sample dominated by Trojans, as opposed to
foreground main-belt asteroids, is provided by the sky-plane surface
density distribution of the 93 identified objects (Figure~3).  This
distribution is peaked towards L4 in a manner incompatible with the
azimuthally uniform distribution of the main-belt asteroids.

\section{Discussion}

\noindent {\it (a) 	Luminosity Function}
	
Photometry was performed using a circular aperture $4.7''$ in projected
diameter, with sky subtraction from a contiguous annulus $3''$ wide
(Table 1).  This aperture was selected by trial and error to give a
stable measure of the flux while minimizing photometric noise from the
background sky.  The statistical photometric uncertainty is $\pm$0.2
mag at the faint end and less than $\pm$0.1 mag at the bright end
of the magnitude distribution.  For comparison with other work, it is
useful to employ the absolute $V$ magnitude, $V(1,1,0) = V - 5\log_{10}(R
\Delta)$, where $R$ [AU] and $\Delta$ [AU] are the heliocentric and geocentric
distances, respectively.  The correction to zero phase angle has been
ignored (since the phase angles near opposition are small).  The
distances $R$ and $\Delta$ are not accurately known for the individual Trojan
asteroids.  We adopt the mean distances of the numbered L4 Trojans at
the epoch of observation, namely $R = 5.1 \pm 0.2$ AU and $\Delta = 4.1 \pm
0.2$ AU.  Here, the quoted errors are $1\sigma$ standard deviations
and the dispersion in $R$ and $\Delta$ results from the finite extent of
the L4 swarm along the line of sight.  With these values we obtain
$V(1,1,0) = V - (6.60 \pm 0.24)$.  For reference, we further compute
the Trojan radii, $r$ [km], from the relation

\begin{equation}
r = \sqrt{\frac{2.24 \times 10^{16+0.4(V_{\rm Sun} - V(1,1,0))}}{p_V}}
\end{equation}

\noindent where $p_V$ is the geometric albedo at the $V$ wavelength
and $V_{\rm Sun} = -26.74$ is the apparent $V$ magnitude of the sun.
The mean geometric albedo of the Trojans recorded by the IRAS
satellite (Tedesco 1989) is $p_V = 0.040 \pm 0.005$ (however, the
quoted statistical uncertainty is probably smaller than inherent
systematic uncertainties due to assumptions made in the radiometric
modelling of the IRAS data).  With this $p_V$, we obtain

\begin{equation}
r_{0.04} {\rm [km]} = 10^{0.2(24.23 - V)}
\end{equation}

\noindent (Figure~4).  A Trojan at the 50\% detection threshhold $V = 22.5$ has
$V(1,1,0) = 15.9$ and $r_{0.04}$ = 2.2 km.  The brightest (faintest)
Trojan detected in the present survey has $V = 17.7 (23.4)$
corresponding approximately to $r_{0.04} = 20.2$ km (1.5 km).  The distance
variation across the diameter of the L4 swarm introduces an
uncertainty to the derived radii of about $\pm 10$\%.

Figure~5 shows the cumulative luminosity function (CLF) computed from
the present data.  The ``$\times$'' marks in the figure show the distribution
of the raw counts while the filled circles show the distribution
corrected for the detection efficiency.  We have not included a
correction for contamination of the Trojan sample by main-belt
interlopers.  As noted above, this is a small effect whose inclusion
would decrease the estimated surface densities at all magnitudes by
about 10\%.  Error bars in Figure~5 were estimated from Poisson
statistics.  The luminosity function is taken to be of the form

\begin{equation}
N(V) {\rm d} V = 10^{\alpha(V-V_0)}~{\rm d} V
\end{equation}
		
\noindent A least-squares fit to the differential magnitude
distribution gives slope parameter $\alpha = 0.40 \pm 0.05$.  The
three lines in Figure~5 have gradients $\alpha-1\sigma, \alpha$, and
$\alpha + 1\sigma$, where $1\sigma = 0.05$.  We assume that the radii
of Trojans follow a differential power law distribution such that the
number of objects having radii in the range $r \rightarrow r+{\rm d}r$ is
$n(r)~{\rm d}r = \Gamma~r^{-q} {\rm d}r$, where $\Gamma$ and $q$ are constants.
For objects all located at a single heliocentric and geocentric
distance, $\alpha$ and $q$ are related by $q = 5\alpha + 1$ (Irwin et
al. 1995).  Thus, the present data suggest $q = 3.0 \pm 0.3$ in the
radius range $2.2 \leq r_{0.04} \leq 20$ km.  For comparison,
Shoemaker et al. 1989 estimated $\alpha = 0.433$ for $10.25 \leq
B(1,0) \leq 14$~mag (corresponding to $q = 3.17$ in the approximate range
$4 \leq r_{0.04} \leq 40$ km, assuming $B-V \sim 0.65$), but did not
state their uncertainty.  We consider these determinations to be in
good agreement.

The difference between $q$ measured here and the canonical $q = 3.5$
distribution produced by collisional shattering (Dohnanyi 1969) is
statistically insignificant.  In any case, unmodeled effects will
cause the Trojan distribution to differ from a Dohnanyi power law.
For example, the velocity of ejection of collision fragments varies
inversely with fragment size (as $r^{-1/2}$, Nakamura and Fujiwara 1991).
Following collisional production the small Trojans should
preferentially escape from the L4 region, leading to a distribution
flatter than the Dohnanyi power law (i.e., $q < 3.5$).  We consider it
likely that the small Trojan asteroids are collisionally produced
fragments of once larger bodies (c.f. Marzari et al. 1997).

\noindent {\it (b) 	Inclination-Frequency Distribution}
	
We are able to measure the position angles of the proper-motion
vectors of the Trojans with an uncertainty of about $\pm 1^{\circ}$.  The
proper-motion vectors cannot be accurately converted to orbital
inclinations without a fuller knowledge of the orbits than we possess.
The approximate relation between the direction of apparent motion and
the true inclination is given by

\begin{equation}
\tan\theta_a = \frac{\tan(i)}{\sqrt{R[1+\tan^2(i)]}-1}
\end{equation}

\noindent where $\theta_a$ is the angle between the apparent direction
of motion and the projected ecliptic, $i$ is the true orbital
inclination, and $R$ is the semi-major axis of the orbit.  In deriving
Eq.~(5) we have assumed that the orbital eccentricities are zero, that
the proper motion is the vector difference of the intrinsic motion of
the Trojans from the orbital motion of earth and that the observations
are taken at opposition.  In the limit $i \rightarrow 90^{\circ}$,
$\tan\theta_a \rightarrow R^{-1/2}$ so that with $R = 5.2$ AU we find
$\theta_a \rightarrow 24^{\circ}$.  One object in Table 1 has $\theta_a >
24^{\circ}$ (\#26).  Presumably, it is a main belt asteroid whose
sky-plane velocity falls fortuitously within the Trojan domain.

The distribution of intrinsic inclinations (Figure~6a) has a mean $i =
7.4^{\circ} \pm 0.7^{\circ}$, median $i = 6.2^{\circ}$.  Trojans with
large inclinations spend most of their time away from the ecliptic,
leading to a bias against their detection.  The bias correction varies
approximately in inverse proportion to the orbital inclination.
Figure~6b shows the inclination distribution after correction by the
factor $1/i$ and normalization to Figure~6a in the range $12 \leq i
\leq 14^{\circ}$.  The bias-corrected mean inclination is $\bar{i} =
13.7^{\circ} \pm 0.5^{\circ}$ which compares favorably with Shoemaker
et al.'s 1989 best estimate for the mean inclination of the larger
Trojans ($\bar{i} = 17.7^{\circ}$).

Observers of Trojan asteroids two decades ago suspected that ``there
is a possibility that the inclinations are bimodal... with groups
separated by a minimum at $i \sim 13^{\circ}$'' (Degewij and van
Houten 1979).  There is indeed an apparent lack of Trojans in the
corrected inclination distribution with $i = 14^{\circ} \pm
1^{\circ}$, giving a bimodal appearance (Fig. 6b).  However,
inspection of the raw data in Figure~6a shows that the local minimum
is statistically insignificant.  Furthermore, the inclination
distribution of numbered and un-numbered L4 Trojans (from an
electronic list maintained by Brian Marsden at the Minor Planet
Center) shows no evidence for bimodality (Figure~7).  Therefore, we
conclude that the data provide no compelling evidence for a bimodal
distribution of inclinations.

\noindent {\it (c) Size and Content of the L4 Trojan Swarm}
	
Figure~3 shows the variation of the surface density, $\Sigma(\theta)$~[deg$^{-2}$], of L4 Trojans with angular distance, $\theta$, from the
L4 point along the ecliptic.  The data can be fitted by a Gaussian
function

\begin{equation}
\Sigma(\theta) d\theta = \Sigma_0 + \Sigma_1\exp \left[\frac{-\theta^2}{2\sigma_T^2} \right] d\theta,~~~~~~~(15^{\circ} \leq \theta \leq 60^{\circ})
\end{equation}

\noindent with $\Sigma_0 = 1.1 \pm 0.4$~[deg$^{-2}$], $\Sigma_1 = 27.7
\pm 4.6$~[deg$^{-2}$] and $\sigma_T = 11.2^{\circ} \pm
0.9^{\circ}$~[deg$^{-2}$].  The apparent FWHM of the L4 swarm measured
along the ecliptic is $26.4^{\circ} \pm 2.1 ^{\circ}$, with measurable
surface density for $\theta \leq 40^{\circ}$.  The linear size
corresponding to FWHM = $26.4^{\circ}$ is $\sim$2.4 AU.  For
comparison, Holman and Wisdom (1992) found that the theoretical stable
zones of the Jupiter Lagrange point have an angular radius of about
$35^{\circ}$.  The angular half width of the swarm along the ecliptic
($13.2^{\circ} \pm 1.0^{\circ}$) is comparable to the bias-corrected
mean inclination ($13.7^{\circ} \pm 0.5^{\circ}$), meaning that we may
take the projected outline of the swarm as approximately circular in
the plane of the sky.
	
The number of L4 Trojans within angle $\theta_{\rm max}$ of L4 is then

\begin{equation}
N(\theta_{\rm max}) = \int^{\theta_{\rm max}}_{0^{\circ}} 2\pi \sin(\theta) \Sigma(\theta){\rm d}\theta.
\end{equation}

\noindent We plot solutions to Eq.~(7) in Figure~8.  The amplitude of
libration about L4 ranges up to $60^{\circ}$ (Shoemaker et al. 1989).
Accordingly, we obtain the nominal population estimate $N(60^{\circ})
= 3.4 \times 10^4$ ($V \leq 22.5, r_{0.04} \geq 2.2$ km), plotted in
Figure 8 as Model A.  The uncertainty on this number may be estimated
in two ways.  Uniformly increasing (decreasing) the fitted parameters
$\Sigma_0$, $\Sigma_1$ and $\sigma_T$ by $1\sigma$ changes $N$ by
$\pm35$\% (c.f. Figure~8).  Systematic errors may also affect
$N(\theta_{\rm max})$, particularly since the surface density at
$\theta \leq 15^{\circ}$ is not measured in our survey.  However, we
find these errors to be small.  If, to consider an extreme case, we
arbitrarily (and unphysically) assume that $\Sigma(\theta \leq
15^{\circ}) = 0$, we obtain (from Eq.~7) $N(60^{\circ}) = 2.1 \times
10^4$, still only 40\% less than the nominal estimate. We conclude
that $N(\theta_{\rm max})$ is uncertain to within a factor of order 2.

The cumulative luminosity function is replotted in Figure~9, including
132 numbered and 125 unnumbered Trojan asteroids (hollow circles)
from orbital element catalogs maintained by the Minor Planet Center.
We used $V(1,1,0) = H + 0.36$ to correct the catalog magnitudes to the $V$
band magnitudes employed here.  In making the comparison between the
number of Trojans deduced from the present survey (Eq.~8) with those
from the Minor Planet lists, we have corrected the former by a factor
of two, to account for the unobserved L5 swarm. (Early suspicions that
L4 might be more populated than L5 have not been borne out by recent data,
supporting our application of a factor of two, Shoemaker et al. 1989).
Curvature of the CLF at $V(1,1,0) \geq 9.5~(r_{0.04} \leq 42$ km) indicates
observational incompleteness in the Minor Planet Trojan sample, as
does the fact that the catalog asteroids are less numerous than those
of the present survey by 1 to 2 orders of magnitude in the common $11
\leq V(1,1,0) \leq 14$ mag range (Figure~9).  Thirty seven
objects have $V(1,1,0) < 9.5$.  Their effective CLF has slope
$\alpha = 0.89 \pm 0.15$, significantly steeper than measured from the
fainter objects of the 8k survey.  The implied size distribution index
is $q = 5.5 \pm 0.9~(V(1,1,0) < 9.5, r_{0.04} \geq 42$ km).

The data of Figure~9 are adequately fitted by the following
differential size distributions:

\begin{equation}
n_1 \left(r_{0.04} \right) {\rm d}r_{0.04} = 1.5 \times 10^6 \left( \frac{1{\rm~km}}{r_{0.04}} \right)^{3.0 \pm 0.3} {\rm d}r_{0.04}~~~~~(2.2 \leq r_{0.04} \leq 20{\rm~km})
\end{equation}

\noindent and 

\begin{equation}
n_2 \left(r_{0.04} \right) {\rm d}r_{0.04} = 3.5 \times 10^9 \left( \frac{1{\rm~km}}
{r_{0.04}} \right)^{5.5 \pm 0.9} {\rm d}r_{0.04}~~~~~(r_{0.04} \geq 42{\rm~km}).
\end{equation}

\noindent The corresponding integral distributions are

\begin{equation}
N(>r_{0.04}) = 1.6 \times 10^5 \left( \frac{1{\rm~km}}{r_{0.04}}
\right)^{2.0 \pm 0.3}~~~~~(2.2 \leq r_{0.04} \leq 20{\rm~km})
\end{equation}

\noindent and

\begin{equation}
N(>r_{0.04}) = 7.8 \times 10^8 \left( \frac{1{\rm~km}}{r_{0.04}} \right)^{4.5 \pm 0.9}~~~~~(r_{0.04} \geq 42{\rm~km})
\end{equation}

\noindent From Eq.~(10) we find the number of L4 Trojans with
$r_{0.04} \geq 5$ km is $N \sim 6400$, to within a factor of order 2.
For comparison, there were 5700 numbered and 1100 unnumbered main-belt
asteroids with $r_{0.04} \geq 5$ km as of 1999 July 29.  These
estimates validate the assertion by Shoemaker et al. (1989) to the
effect that the populations of the main-belt and the Trojan swarms are
of the same order.  The number of L4 Trojans with $r_{0.04} \geq 1$ km
is $\sim 1.6 \times 10^5$, from Eq. (11).
		
The most straightforward explanation of the slope differences (Eq.~8
and 9 and Figure~9) is that the large objects represent a primordial
population while Trojans smaller than a critical radius, $r_c$, are
produced from the larger ones by collisional shattering (Shoemaker et
al. 1989, Marzari et al. 1996).  By equating Eqs.~(10) and (11) we
find $r_c \sim 30$ km (corresponding to $V(1,1,0) = 10.2 \pm 0.5$
(Figure~9)).  Binzel and Sauter (1992) reported that Trojans with
$r_{0.04} > 45$ km have a larger mean lightcurve amplitude (a measure
of elongated body shape) than their low albedo main-belt counterparts.
They suggested that this might mark the primordial/fragment transition
size, with larger bodies retaining the aspherical forms in which they
were created (we note that this explanation is clearly non-unique).
Inspection of their Figure~21 shows that the transition radius defined
in this way is uncertain to within a factor of 2 and fully compatible
with $r_c \sim 30$ km as found here.  We conclude that two
independently measured physical parameters (the size distribution and
the lightcurve amplitude distribution) show evidence for a change near
$r_c \approx 30$ km to 40 km.

The total mass of Trojans is

\begin{equation}
M_T = \int^{r_c}_0 \frac{4}{3} \pi \rho n_1(r){\rm d}r + \int^{\infty}_{r_c} \frac{4}{3} \pi \rho n_2(r) {\rm d}r
\end{equation}

\noindent where $n_1$ and $n_2$ are from Eqs.~(8) and (9) and $\rho =
2000$ kg m$^{-3}$ is the assumed bulk density.  We find $M_T \approx 5
\times 10^{20}$ kg $\approx 9 \times 10^{-5} M_{\rm Earth}$, equivalent to
a 400 km radius sphere having the same density. Dynamical calculations
show that escaped Trojans would be quickly scattered into orbits
indistinguishable from those of some short-period comets (Marzari et
al. 1995).  Therefore it is of interest to compare $M_T$ with the mass of
short-period comets delivered to the inner solar system over the past
4.5 Gyr.  The rate of supply of short period comets is $f \sim
10^{-2}$ yr$^{-1}$ (Fernandez 1985).  The size and mass distributions
of the cometary nuclei have not been adequately measured.  We assume
that the nuclei follow a differential size distribution with index $q
= 3$ and minimum and maximum radii $r_1 = 0.5$ and $r_2 = 30$ km.  The
mass-weighted mean radius drawn from this distribution is $\bar{r}
\approx (2r_2 r_1^2)^{1/3} = 2.5$ km.  A small number of well-measured
cometary nuclei have radii $\sim$ few km (Jewitt 1996), consistent
with this estimate.  The delivered mass of short-period comets is then
$M_C \approx 4 \pi \rho \bar{r}^3 fT/3$, where $T = 4.5 \times 10^9$
yr is the age of the solar system.  We find $M_C \approx 6 \times
10^{21}$ kg, corresponding to $M_C \approx 10~M_T$.  This mass, if taken
at face value, makes it unlikely that the
Trojan swarms could be the dominant source of the comets.  It is
entirely possible, however, that a fraction of the short-period comets
could be escaped Trojans. A definitive estimate of this fraction will require better knowledge of the cometary parameters ($r_1$, $r_2$, $q$, $f$, and 
density) than we now possess, as well as detailed understanding of the physics
of collisional ejection from the Trojan swarms.

The present results were extracted from data taken for an independent
(Kuiper Belt) purpose.  They serve to give an idea of the power of
modern CCD arrays on a telescope of rather modest diameter.  Much more
could be learned from a survey specifically targeting the Trojan asteroids and
including astrometric follow-up, so that orbital elements can be
determined for individual objects.  Future work should focus on a more
complete digital survey of both L4 and L5 swarms in order to determine
the total population and size distribution with greater confidence.
Observations taken away from the ecliptic will provide a better
measure of the high inclination objects.  Observations to fainter
limiting magnitudes will allow us to probe the sub-kilometer
population.  Carefully planned observations will produce stronger
constraints on the collisional and dynamical states of the Jovian
Trojans, leading ultimately to a deeper understanding of these enigmatic bodies.

\section{Summary}

\begin{enumerate}

\item The luminosity function of the Jovian L4 Trojans has slope of
$0.40 \pm 0.05$ in the magnitude range $18.0 \leq V \leq 22.5$,
corresponding to objects with radii $2 \leq r_{0.04} \leq 20$~ km (where
$r_{0.04}$ is the radius derived assuming a geometric albedo of 0.04).
The corresponding differential power law size distribution index is $q
= 3.0 \pm 0.3$.  This is consistent with the slope expected for a
collisionally shattered population ($q \sim 3.5$; Dohnanyi 1969)
within $\sim 2\sigma$ (95\%) confidence and suggests that the small
Trojans are collisional fragments of larger bodies.

\item The brighter (larger) Trojans follow a $q = 5.5 \pm 0.9$
differential power law distribution.

\item The apparent FWHM of the L4 swarm is $26^{\circ} \pm 2^{\circ}$, measured
along the ecliptic.

\item The distribution of inclinations of the Trojans, when corrected
for observational bias, has a mean $13.7^{\circ} \pm 0.5^{\circ}$.

\item About $1.6 \times 10^5$
L4 Trojans are bigger than 1 km radius.  Their combined mass is of
order $5 \times 10^{20}$ kg ($9 \times 10^{-5} M_{\rm Earth}$), assuming
bulk density $\rho = 2000$ kg m$^{-3}$.

\end{enumerate}

\noindent {\bf Acknowledgements}

\noindent We thank John Dvorak for operation of the UH
telescope. Gerry Luppino, Mark Metzger, Richard Wainscoat and Pui Hin
Rhoads helped us with the camera and the computer set-up.  We
benefitted from digital lists of Trojan orbital elements maintained by
Brian Marsden and David Tholen. Scott Sheppard and an anonymous
referee provided helpful comments. This research was supported by
grants from NASA to DCJ.

\newpage

\centerline{\bf FIGURE CAPTIONS}

\noindent {\bf Figure~1}. The sky coverage of the present survey is
marked by squares.  The L4 Lagrangian point is marked by a large
filled circle and the ecliptic is shown for reference.

\noindent {\bf Figure~2}.  (a) Apparent angular velocities of numbered
main-belt asteroids (empty circles) and Trojans (filled circles)
projected in the direction of observation on 1996 October 10.  Marked
curves show angular velocities $\omega = 22''$/hr and 24$''$/hr.  (b) Objects
detected by MODS having $\omega \leq 24''$/hr: those with $\omega \leq 22''$/hr
are taken to be Trojans for the purpose of this study.  All objects
moving faster than $\omega = 24''$/hr are main-belt asteroids and were
ignored by the MODS software.

\noindent {\bf Figure~3}.  Surface density variation of the observed L4
swarm along the ecliptic.  A Gaussian distribution with $\sigma_T =
11.2 \pm 0.9$ (corresponding to FWHM is $26.4^{\circ} \pm
2.1^{\circ}$) fits the data.

\noindent {\bf Figure~4}.  Magnitude vs. size relation for Trojan
asteroids (Eq.~2).

\noindent {\bf Figure~5}.  Cumulative magnitude-frequency distribution
of detected Trojans.  Crosses mark the original data; filled circles
indicate the distribution after correction for the detection
efficiency.  The straight lines have gradients $\alpha$ = 0.35, 0.40 and
0.45.

\noindent {\bf Figure~6}.  (a) Apparent distribution of inclinations of
Trojans found in the present survey.  (b) Same as (a) but corrected for
inclination bias and normalized at $12^{\circ} \leq i \leq 14^{\circ}$.

\noindent {\bf Figure~7}.  Inclination distribution of the L4 Trojans
(numbered + unnumbered).

\noindent {\bf Figure~8}.  Cumulative number of Trojans with radii $\geq
2.2$ km (the smallest objects detected in the present survey) as a function
of angular distance from the L4 point.  Model A shows Eq.~(7) with
$\Sigma_0 = 1.1 \pm 0.4$, $\Sigma_1 = 27.7 \pm 4.6$ and $\sigma_T =
11.2 \pm 0.9$ in the range $0^{\circ} \leq \theta \leq 60^{\circ}$.
Model B is the same as Model A except that $\Sigma_0 = \Sigma_1 \equiv
0$ for $\theta \leq 15^{\circ}$.  Uncertainties on both curves show
the effect of forcing $\Sigma_0$, $\Sigma_1$ and $\sigma_T$ to
$+1\sigma$ and $-1\sigma$ from the best-fit values.

\noindent {\bf Figure~9}.  Cumulative luminosity function from this
work (filled circles) and from 257 cataloged Trojans detected in
earlier surveys (empty circles).  Counts from the present survey have
been doubled to account for the unobserved L5 swarm.

\newpage

%\makeatletter
%\def\jnl@aj{AJ}
%\ifx\revtex@jnl\jnl@aj\let\tablebreak=\nl\fi
%\makeatother

\begin{deluxetable}{rrrrcrrrr}
\footnotesize
\tablecaption{Trojan Asteroids from the 8k Survey}
\tablewidth{0pc}
\tablehead{
\colhead{$N$}&\colhead{$V$}& 
\colhead{$D$ [km]}&\colhead{$i$ [deg]}& \colhead{}&
\colhead{$N$}&\colhead{$V$}& \colhead{$D$ [km]} & \colhead{$i$ [deg]}
}
\startdata
1 & 17.7 & 40.9	& 22.8 & & 48 & 21.6 & 6.7 & 1.5\\
2 & 17.8 & 40.1	& 2.2 & & 49 & 21.6 & 6.8 & 1.8\\
3 & 18.7 & 26.4	& 9.6 & & 50 & 21.7 & 6.4 & 10.8\\
4 & 18.8 & 25.2	& 2.1 & & 51 & 21.8 & 6.2 & 0.6\\
5 & 18.9 & 23.5	& 12.1 & & 52 & 21.8 & 6.1 & 3.8\\
6 & 19.0 & 22.8	& 5.4 & & 53 & 21.8 & 6.2 & 4.5\\
7 & 19.5 & 18.2	& 0.3 & & 54 & 21.8 & 6.4 & 0.7\\
8 & 19.7 & 16.0	& 6.5 & & 55 & 21.8 & 6.3 & 8.3\\
9 & 19.7 & 16.1	& 6.2 &	& 56 & 21.9 & 6.1 & 3.2\\
10 & 19.8 & 15.8 & 8.7 & & 57 & 21.9 & 5.9 & 5.6\\
11 & 19.8 & 15.8 & 6.7 & & 58 & 21.9 & 6.0 & 5.4\\
12 & 20.0 & 14.2 & 7.7 & & 59 & 21.9 & 6.0 & 1.6\\
13 & 20.1 & 13.9 & 15.9	& & 60 & 21.9 & 5.9 & 3.7\\
14 & 20.1 & 13.7 & 5.0 & & 61 & 21.9 & 5.9 & 5.2\\
15 & 20.4 & 11.7 & 20.0	& & 62 & 21.9 & 5.9 & 20.7\\
16 & 20.5 & 11.3 & 16.1	& & 63 & 22.0 & 5.7 & 0.8\\
17 & 20.6 & 10.8 & 8.3 & & 64 & 22.0 & 5.7 & 2.6\\
18 & 20.6 & 10.9 & 8.5 & & 65 & 22.0 & 5.7 & 16.7\\
19 & 20.6 & 10.6 & 9.1 & & 66 & 22.1 & 5.4 & 7.4\\
20 & 20.7 & 10.2 & 7.2 & & 67 & 22.2 & 5.1 & 7.6\\
21 & 20.7 & 10.4 & 0.4 & & 68 & 22.2 & 5.2 & 7.9\\
22 & 20.8 & 9.6	& 1.7 & & 69 & 22.2 & 5.3 & 17.7\\
23 & 20.8 & 9.9	& 8.0 & & 70 & 22.2 & 5.2 & 4.1\\
24 & 20.8 & 10.0 & 7.6 & & 71 & 22.3 & 5.1 & 5.7\\
25 & 20.8 & 9.9	& 0.3 & & 72 & 22.3 & 4.9 & 6.2\\
26 & 20.9 & 9.2	& 29.2 & & 73 & 22.3 & 4.9 & 4.8\\
27 & 20.9 & 9.6	& 12.8 & & 74 & 22.3 & 4.8 & 25.7\\
28 & 20.9 & 9.3	& 8.7 & & 75 & 22.4 & 4.7 & 20.0\\
29 & 20.9 & 9.3	& 9.3 & & 76 & 22.4 & 4.7 & 7.3\\
30 & 21.0 & 8.9	& 5.7 & & 77 & 22.4 & 4.7 & 2.4\\
31 & 21.0 & 9.0	& 1.8 & & 78 & 22.4 & 4.7 & 1.0\\
32 & 21.1 & 8.7	& 2.2 & & 79 & 22.5 & 4.5 & 6.4\\
33 & 21.1 & 8.6 & 8.9 & & 80 & 22.6 & 4.3 & 0.6\\
34 & 21.1 & 8.6	& 18.8 & & 81 & 22.6 & 4.4 & 3.5\\
35 & 21.1 & 8.7	& 8.0 & & 82 & 22.6 & 4.4 & 11.9\\
36 & 21.1 & 8.6	& 7.2 & & 83 & 22.7 & 4.2 & 2.4\\
37 & 21.3 & 7.7	& 3.7 & & 84 & 22.7 & 4.1 & 1.2\\
38 & 21.3 & 7.8	& 3.4 & & 85 & 22.8 & 4.0 & 0.5\\
39 & 21.4 & 7.4	& 6.0 & & 86 & 22.8 & 4.0 & 2.0\\
40 & 21.5 & 7.1	& 10.7 & & 87 & 22.9 & 3.7 & 2.4\\
41 & 21.5 & 7.2	& 1.2 & & 88 & 22.9 & 3.8 & 19.4\\
42 & 21.5 & 7.3	& 10.2 & & 89 & 23.0 & 3.5 & 3.8\\
43 & 21.5 & 7.1	& 3.6 & & 90 & 23.1 & 3.4 & 9.5\\
44 & 21.5 & 7.2	& 8.8 & & 91 & 23.2 & 3.3 & 0.1\\
45 & 21.6 & 7.0	& 2.1 & & 92 & 23.3 & 3.1 & 1.9\\
46 & 21.6 & 7.0	& 16.5 & & 93 & 23.4 & 2.9 & 21.5\\
47 & 21.6 & 6.8	& 6.2 & & & & & \\
\enddata
\end{deluxetable}


\newpage

\begin{references}
{
\reference{} Binzel, R. P. and Sauter, L. M. 1992. Icarus {\bf 95},
pp. 222-238.

\reference{} Chen, J., Jewitt, D., Trujillo, C., and Luu, J. 1997.
Bull. Am. Astr. Soc {\bf 29}, 25.08.

\reference{} Cruikshank, D. 1977.  Icarus {\bf 30}, 224-230.

\reference{} Degewij, J., and van Houten, C. J. 1979.  In {\it
Asteroids}, ed. T. Gehrels (Univisity of Arizona Press, Tucson),
pp. 417-435.

\reference{} Dohnanyi, J. 1969. J. Geophys. Res. {\bf 74}, pp. 2531-2554.

\reference{} Dumas, C., Owen, T., and Barucci, M. 1998.  Icarus {\bf
133}, 221-232.

\reference{} Fernandez, J. A. 1985. Icarus {\bf} 64, pp. 308-319.

\reference{} Fitzsimmons, A., Dahlgren, M., Lagerkvist, C.-I.,
Magnusson, P., and Williams, I.P. 1994.  Astron. Ap. {\bf 282},
634-642

\reference{} Holman, M., and Wisdom, J. 1993. A. J. {\bf 105}, pp.1987-1999.

\reference{} Irwin, M., Tremaine, S., and Zytkow, A. 1995.  A. J. {\bf
110}, 3082.

\reference{} Jewitt, D. C. and Luu, J. X. 1990.  A. J. {\bf 100}, 933-944.

\reference{} Jewitt, D. 1996.  Earth, Moon and Planets {\bf 72}, 185-201.

\reference{} Jewitt, D., Luu, J. and Trujillo, C. 1998.  A. J. {\bf
115}, 2125-2135.

\reference{} Landolt, A. 1992.  A. J. {\bf 104}, 340 - 371.

\reference{} Levison, H. F., Shoemaker, E. M. and Shoemaker,
C. S. 1997. Nature {\bf 385}, pp.42-44.

\reference{} Marzari, F., Farinella, P., and Vanzani, V. 1995.
Astron. Ap. {\bf 299}, 267-276.

\reference{} Marzari, F., Scholl, H., and Farinella, P. 1996. Icarus
{\bf 119}, pp. 192-201.

\reference{} Marzari, F., Farinella, P., Davis, D. R., Scholl, H. and
Bagatin, A. C. 1997. Icarus {\bf 125}, pp. 39-49.

\reference{} Marzari, F., and Scholl, H. 1998.  Astron. Ap. {\bf 339}, 278-285.

\reference{} Nakamura, A., and Fujiwara, A. 1991.  Icarus {\bf 92}, 132-146.

\reference{} Peale, S. 1993.  Icarus {\bf 106}, 308.

\reference{} Shoemaker, E. M., Shoemaker, C. S. and Wolfe,
R. F. 1989. In {\it Asteroids II}, eds. Binzel, R.P., Gehrels, T. and
Matthews, M. S. (University of Arizona Press, Tucson), pp. 487-523.

\reference{} Tedesco, E. F. 1989. In {\it Asteroids II}, eds. Binzel,
R.P., Gehrels, T. and Matthews, M. S. (University of Arizona Press,
Tucson), pp. 1090-1138.

\reference{} Trujillo, C. and Jewitt, D. 1998. A. J. {\bf 115}, 1680-1687.
}
\end{references}
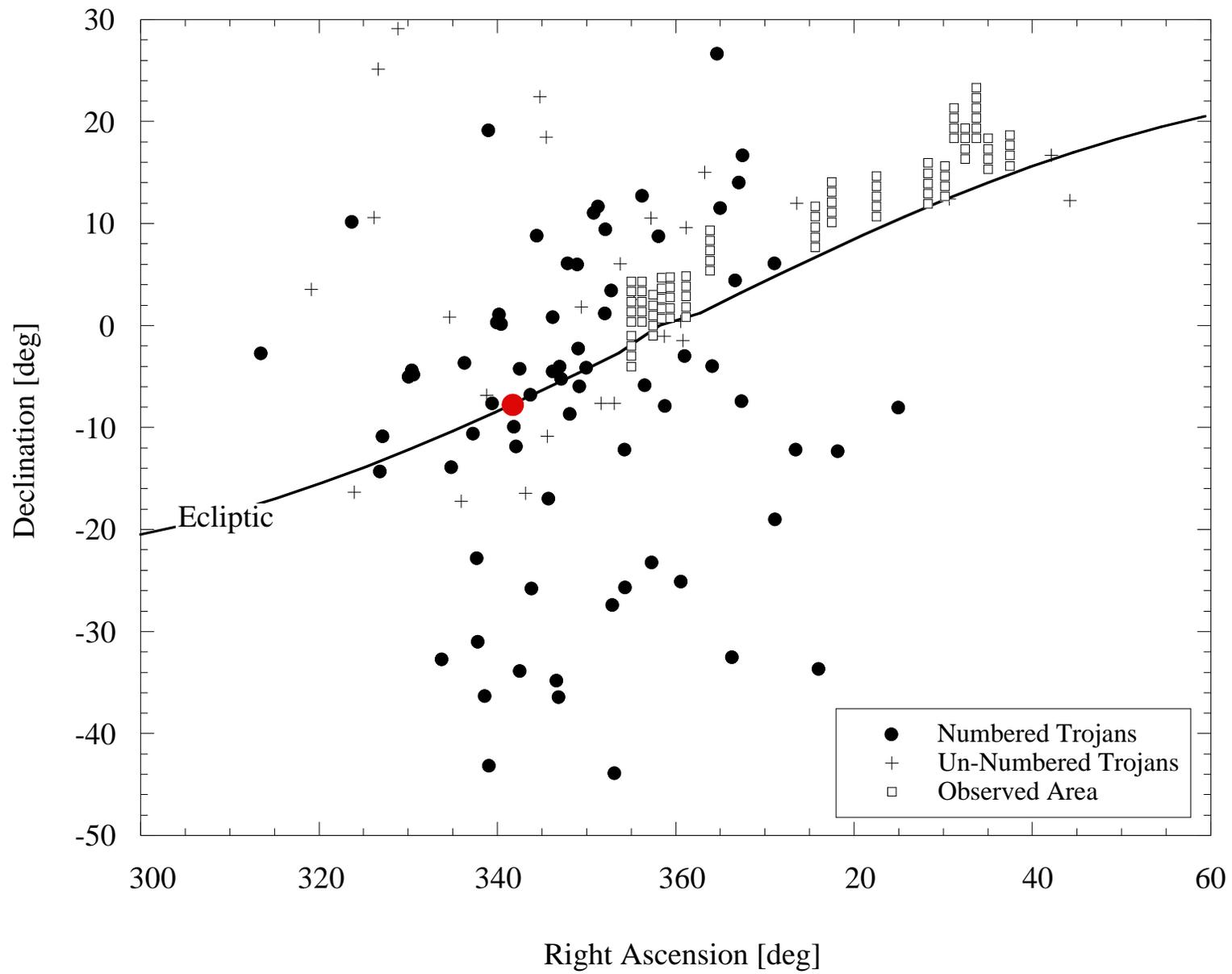
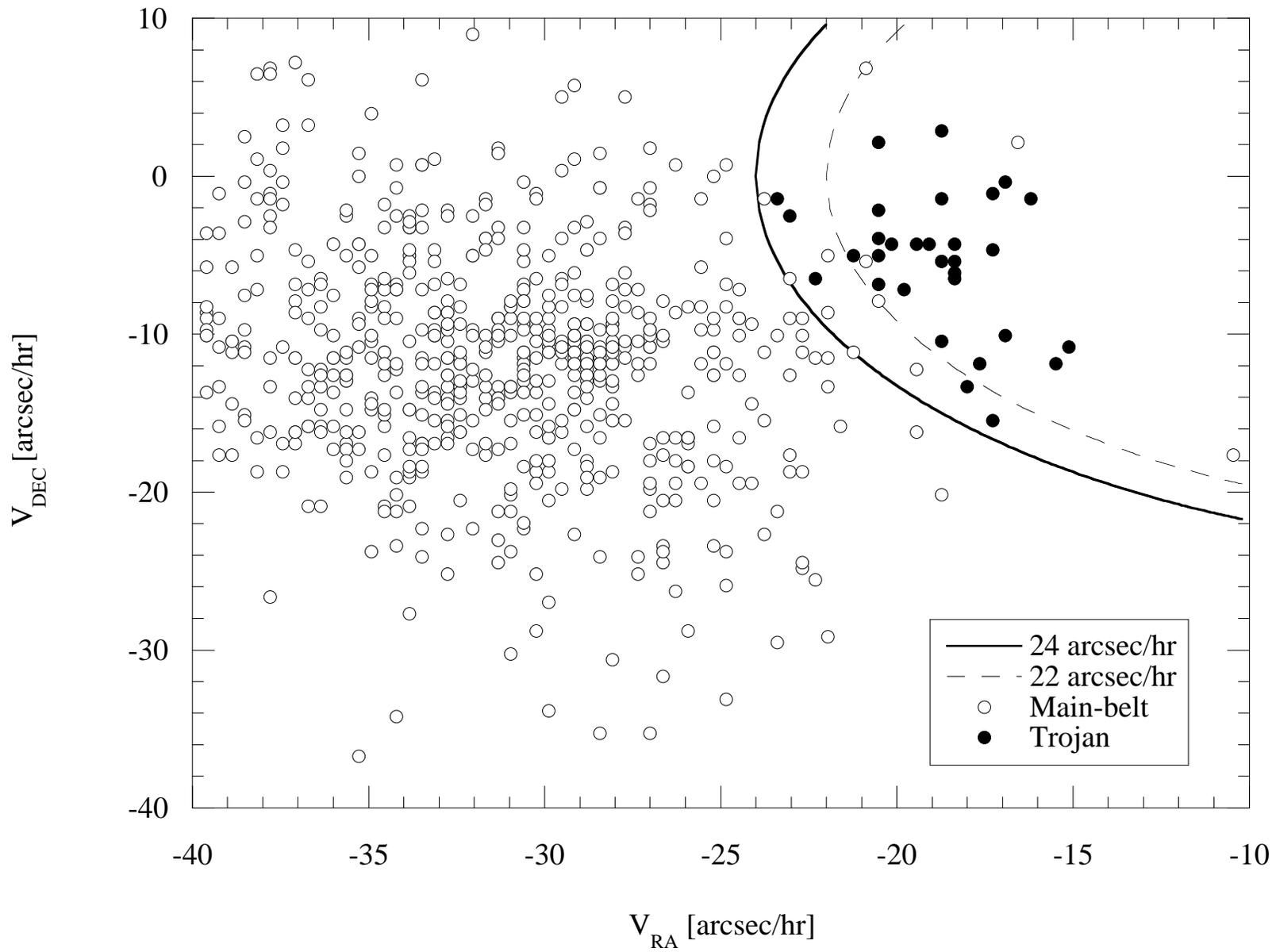
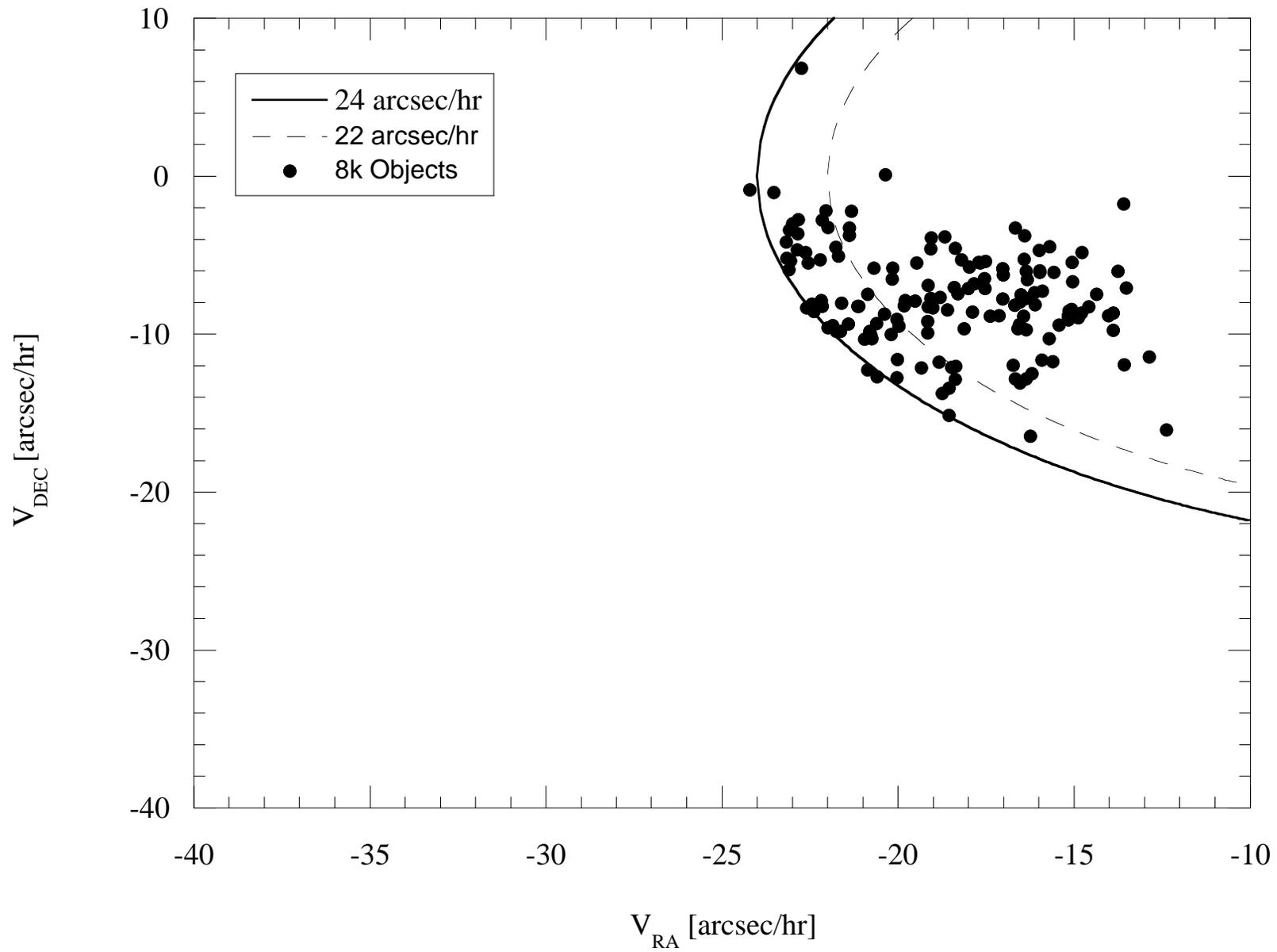
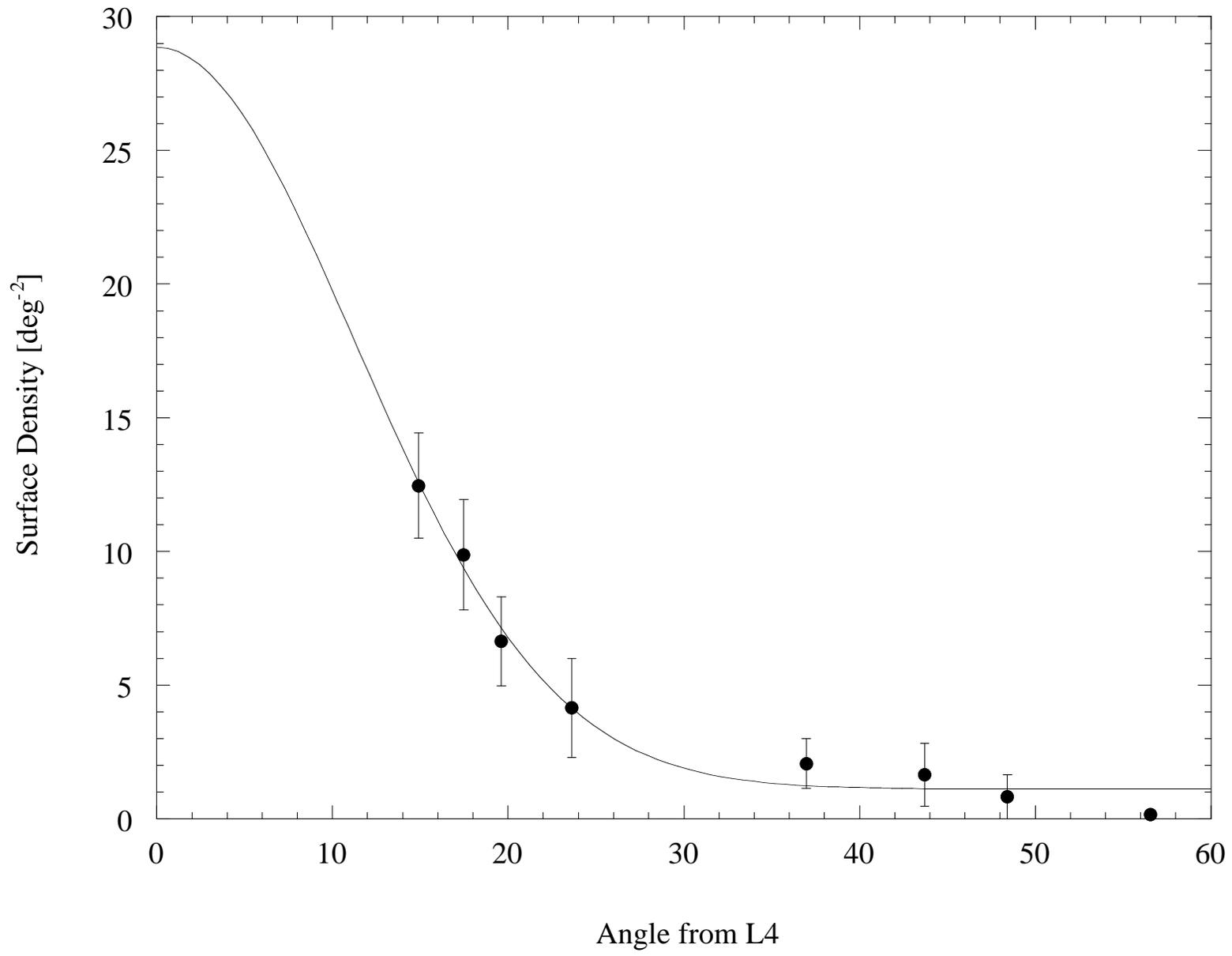
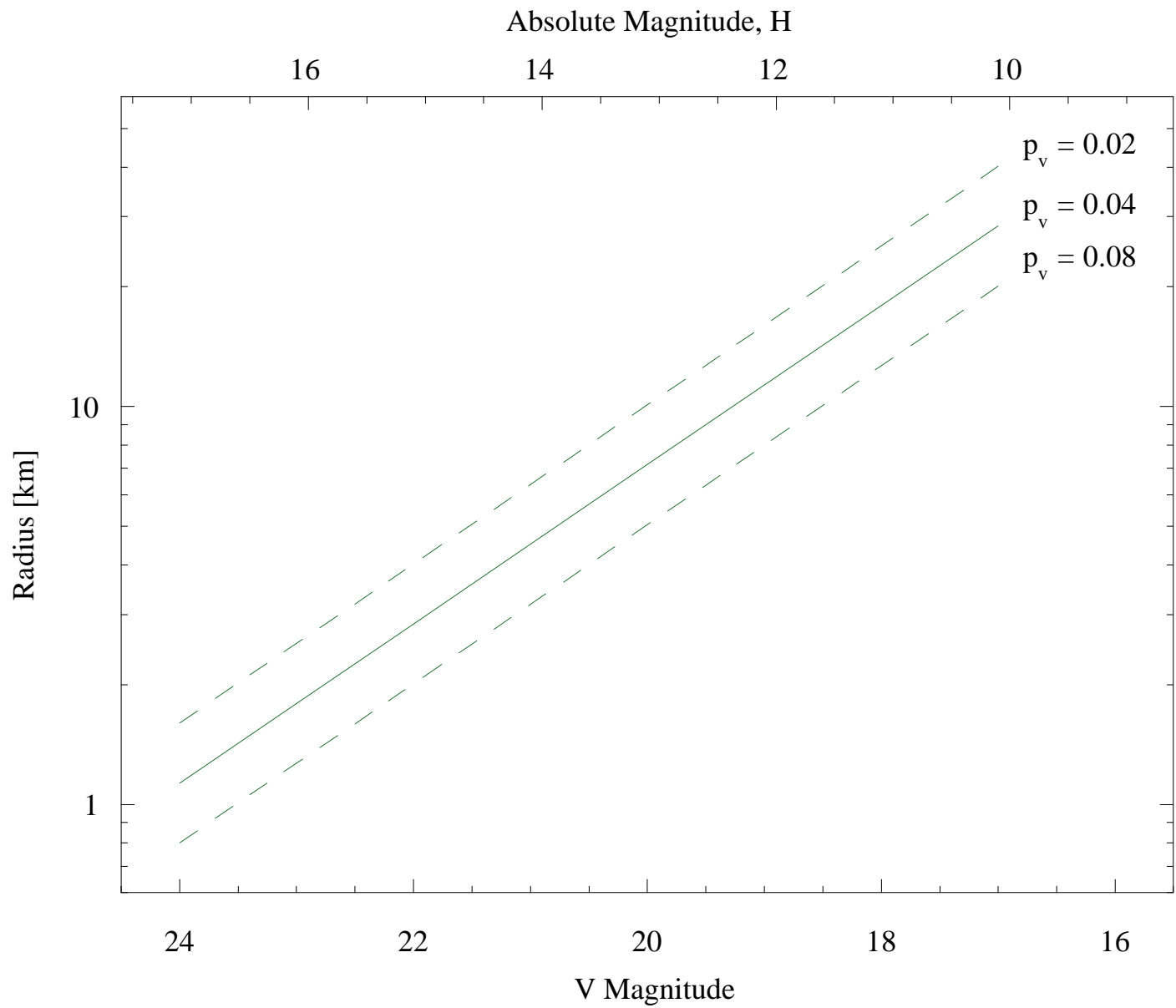
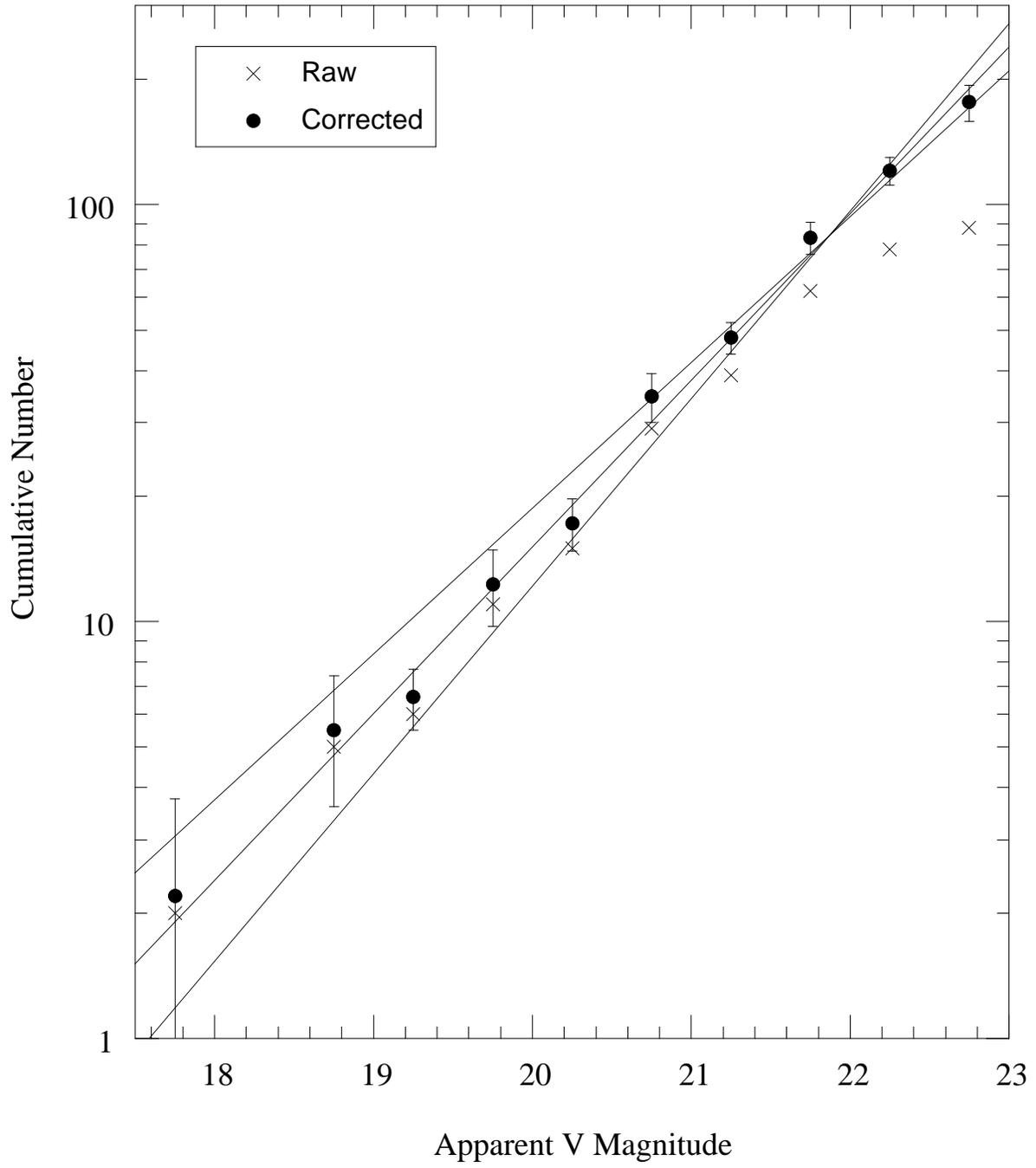
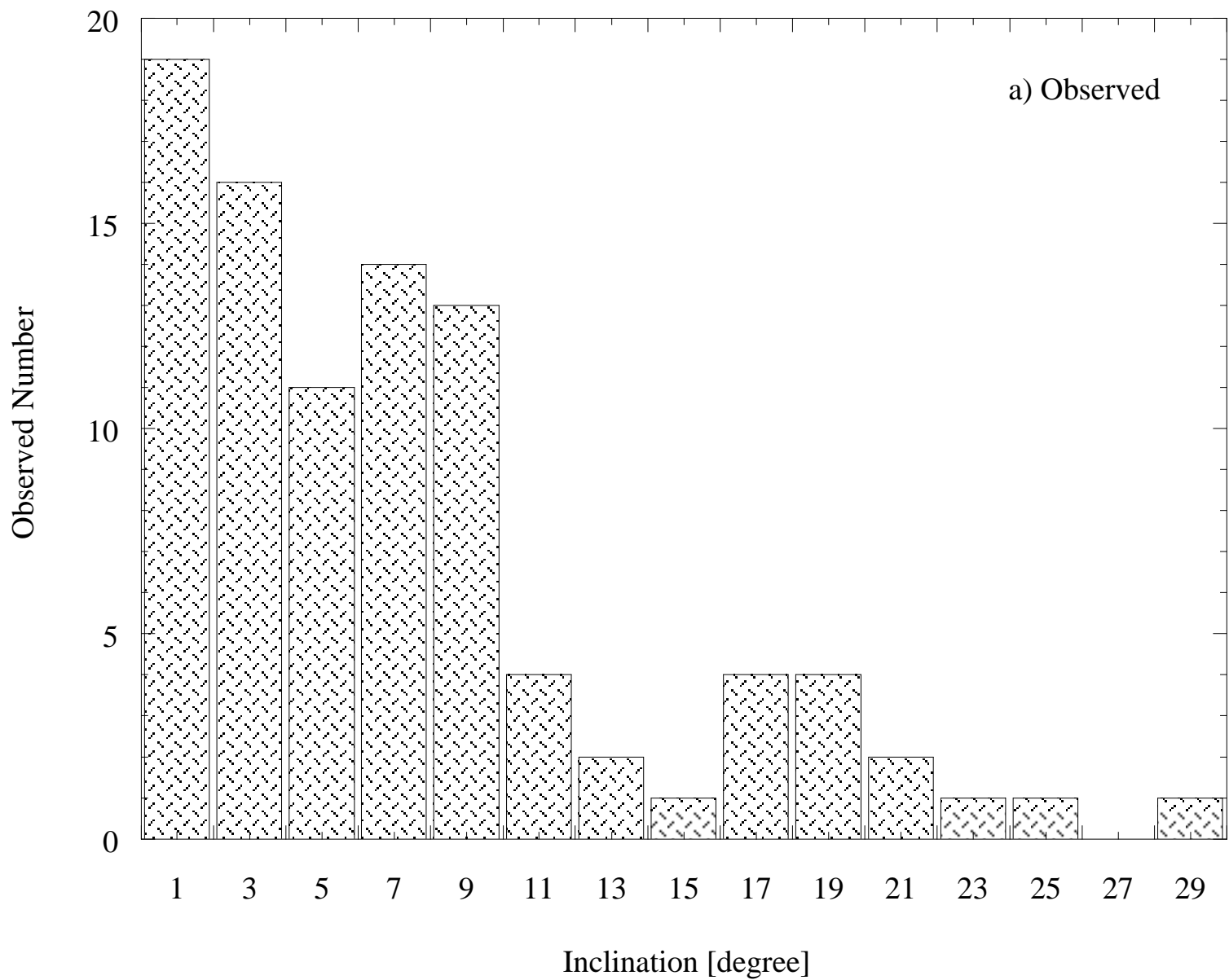
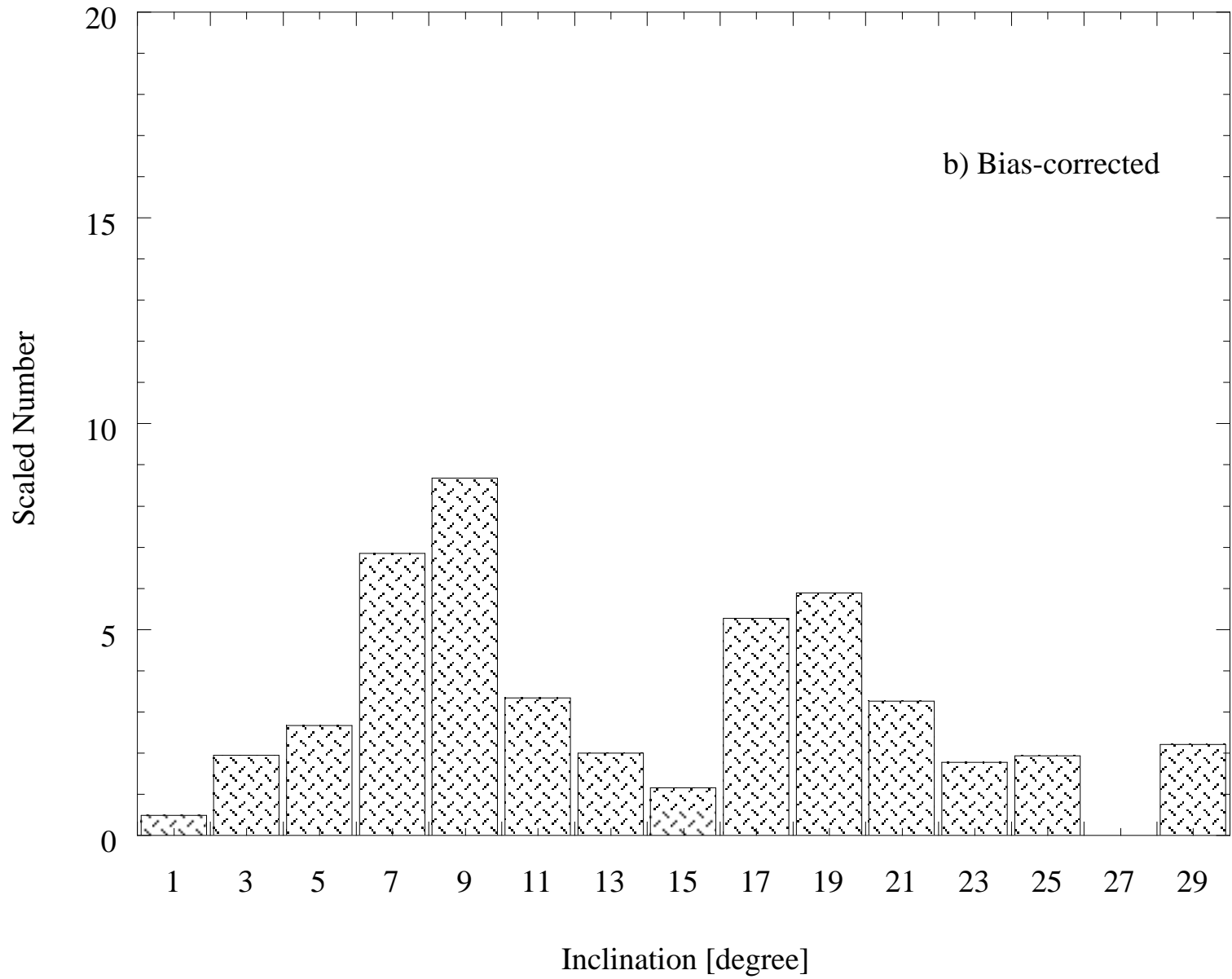
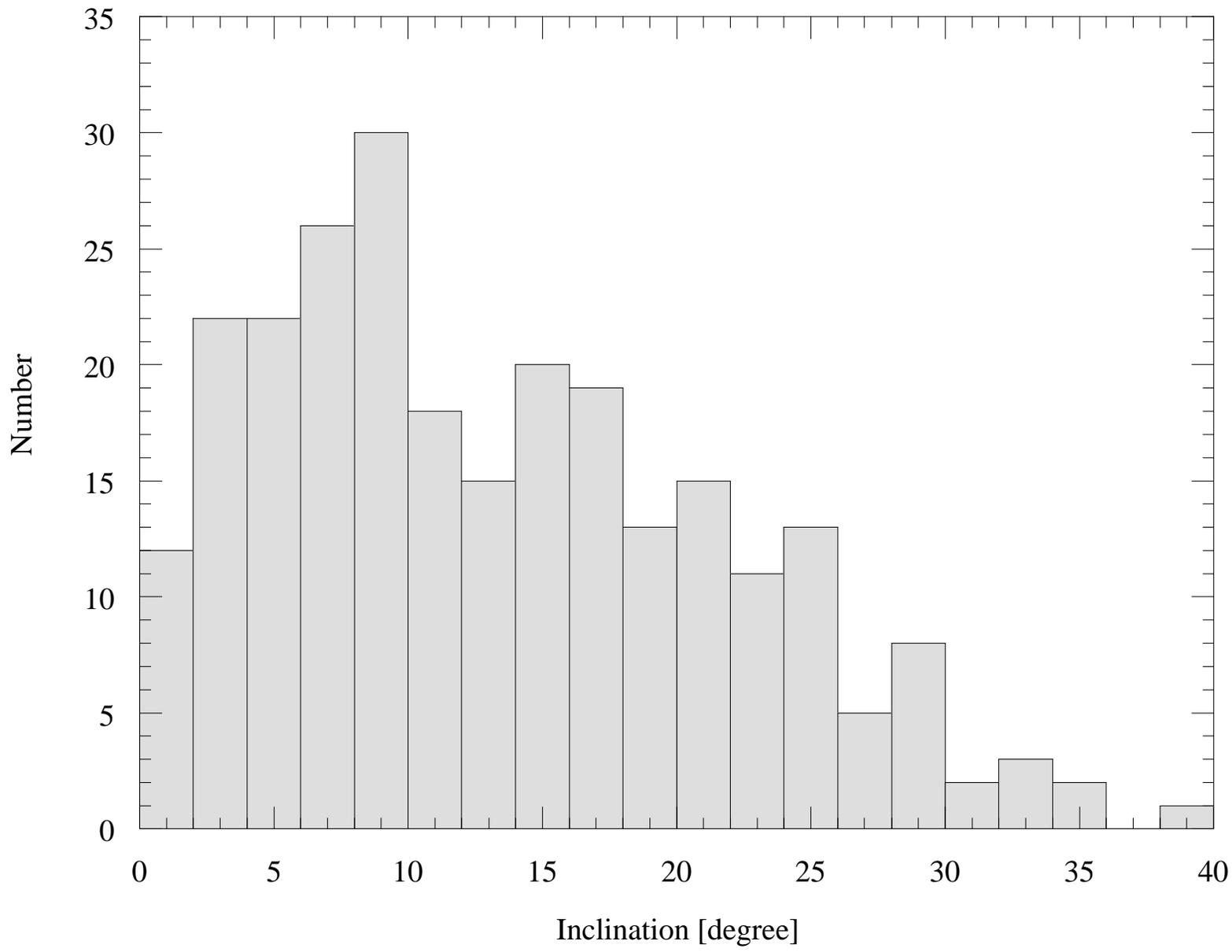
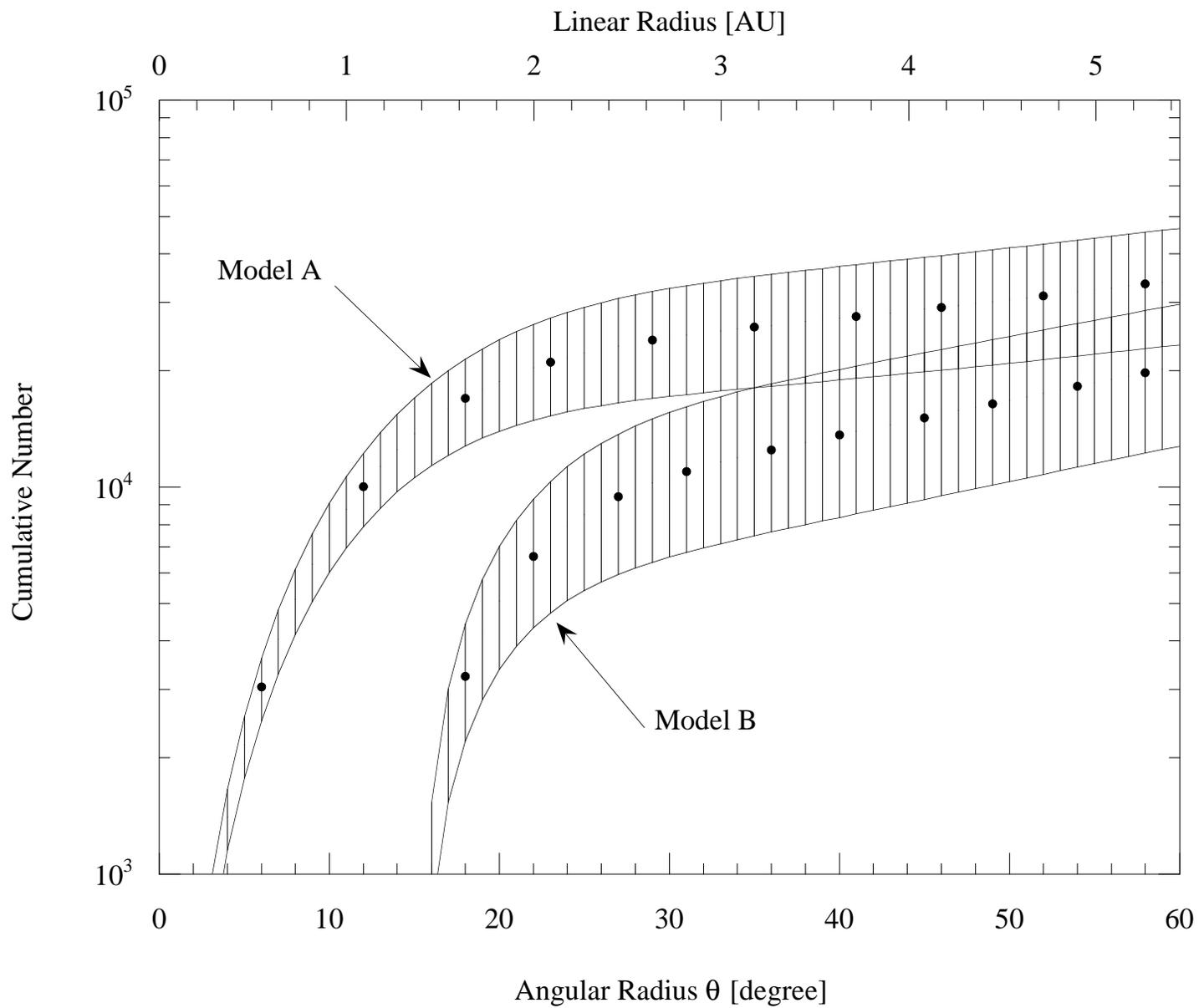
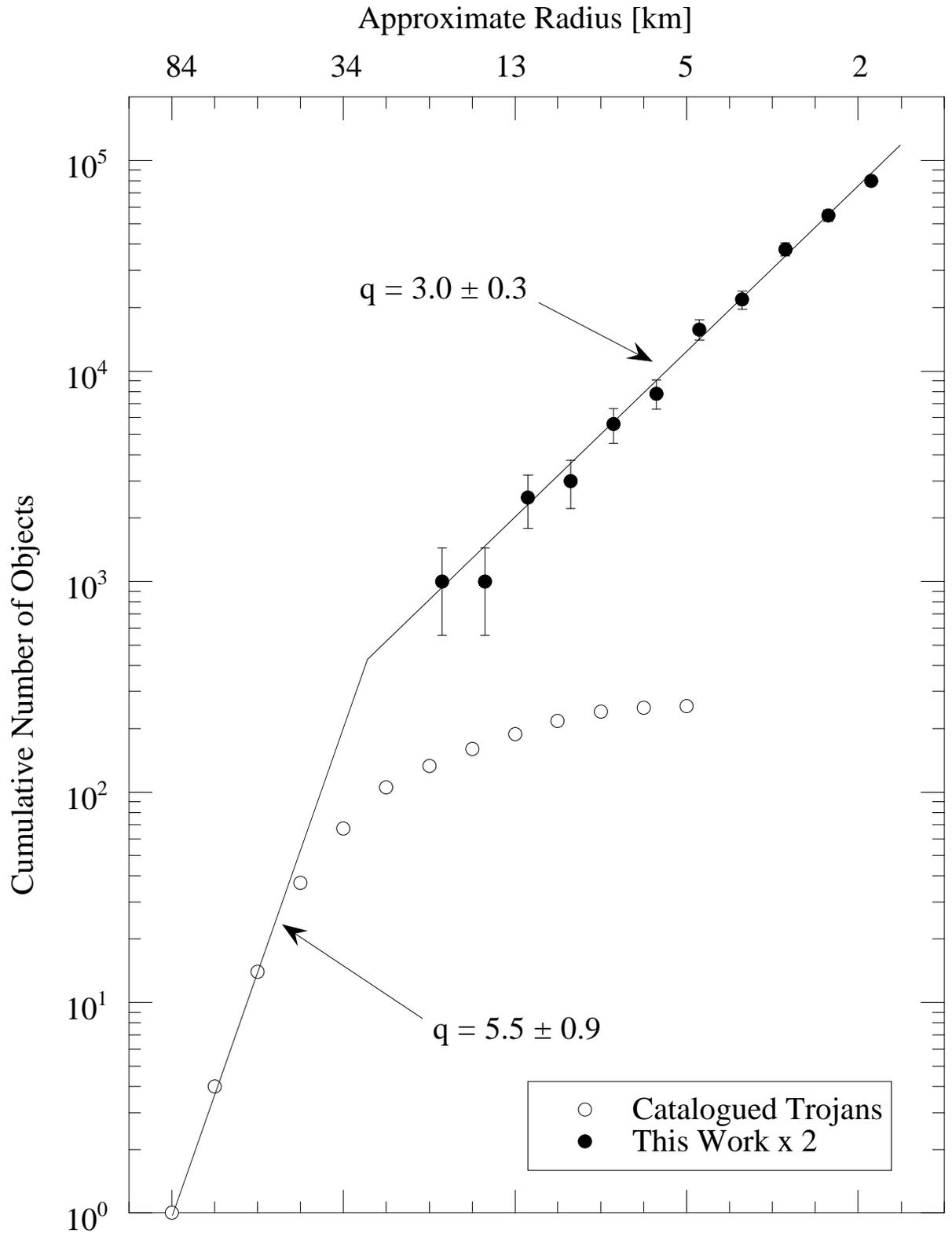
\end{document}